# Complete Performance Analysis of Underwater VLC Diffusion Adaptive Networks

Hossein Abdavinejad[1], Hadi Baghali[1], Javad Ostadieh[1,2], Ehsan Mostafapour[1*],
Changiz Ghobadi[1], Javad Nourinia[1].

[1]*Dept. Electrical Engineering, Urmia University, Urmia, Iran.*
[2] *Dept. Electrical Engineering, Islamic Azad University, Khoy branch, Khoy, Iran.*
Corresponding Author: *e.mostafapour@urmia.ac.ir*

*Abstract*— **In this paper, we simulated a diffusion adaptive network in the underwater environment. The communication method between the nodes of this network is assumed to be the visible light communication technology (VLC) which in the underwater condition is known as the UVLC. The links between the nodes in this case are contaminated with the optical noise and turbulence. These contaminations are modeled with the proper statistical distributions depending on the underwater conditions. The optical turbulence modeling link coefficients are shown to be following the Log-normal distribution which its mean and variance are directly dependent on the temperature and the salinity of the simulated water and the assumed distance between the diffusion network nodes. The performance of the diffusion network in using UVLC technology is then analyzed both with simulations and theoretical calculations and the results are presented using the steady-state error metrics. Our analysis showed that the diffusion network can be implemented underwater with the VLC technology providing that the distance between the network nodes is less than 10 meters. Also, in order to guarantee the convergence of the adaptive network, the water salinity level and temperature must not exceed the values that are presented in our simulations.**

*Index Terms*- Diffusion adaptation; Visible light communication; Underwater; Optical turbulence; Log-normal distribution; Convergence;   Steady-State.

I. INTRODUCTION

There are many useful applications of underwater wireless sensor networks [1]. Multi-wireless underwater technologies such as radio frequency (RF), acoustic and optical signaling are considered as connectivity solutions for USNs. RF which is used for data rates up to 100 Mb/s at close distances is not always preferable since it is expensive and bulky. Acoustic signaling, as another wireless technology, was considered the most effective wireless communication technology in an underwater environment. This is due to its ability of long-distance transmission. On the other hands, acoustic suffer from low propagation speed and small available bandwidth. This raises the demand on third wireless underwater technology, i.e., underwater optical wireless communication (UOWC) [2]. When light travels through the water, the ultraviolet and infrared signals are absorbed first, leaving visible light essentially bound to the blue-green portion of the spectrum as the best wavelengths of underwater transmission [3, Chapter 4]. The green and the blue parts of the visible spectrum have less attenuation in coastal waters and open ocean, respectively. [4, Chapter 8]. In this portion of the spectrum, there is mounting literature on underwater visible light communication (UVLC) [5-8]. However, most of the works are limited to single-user cases and point-to-point communication links. Practical implementation of USNs requires the design of adaptive with multiple access networks for supporting several sensor nodes that can locate and track objects, estimate various parameters and monitor different values using adaptive distributed algorithms [9-12, 27]. Furthermore, diffusion technologies have a huge potential in various applications [12-14] and they are, generally, implemented by using one of two main strategies: The Adapt Then Combine (ATC) and the Combine Then Adapt (CTA).

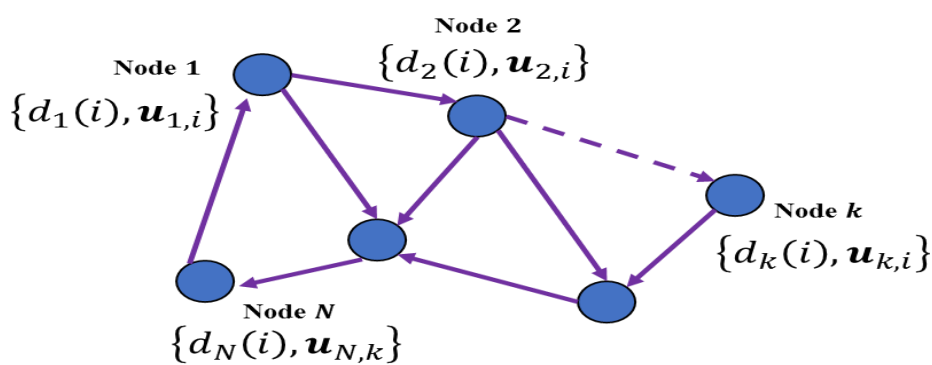

Fig. 1. Adaptive diffusion network.

Each of these strategies is suitable for special applications. In [13], it was that CTA outperforms ATC algorithm in turbulent environments. In underwater, the refractive index fluctuates underwater due salinity and / or temperature fluctuations. This results in underwater turbulence, which in turns fluctuate the received signal over its average that is known as fading. Turbulence induced fading is one of the most impairments of using UVLC [15-19]. The statistical distribution of underwater fading was experimentally investigated. It has been shown, for weak turbulence, that statistical distribution of received intensity follow Log-normal probability density functions (PDF) [27] Recently, the effects of turbulence on the performance of the adaptive diffusion and incremental networks has been investigated in the context of free space optical communications (FSO) in [13, 20-22, 28-29]. However, UVLC with different path loss model and turbulence characteristics may have different performance.

In this paper, based on the findings in [15] and [19], the authors derived the exact parameters for the underwater Log-normal turbulence models and analyzed the exact performance of the diffusion adaptive network when implemented using the underwater VLC technology. If we want to implement adaptive networks underwater with the VLC technology, we must know exactly about the distribution model that best describes the underwater conditions for the VLC technology and the related parameters. For this reason, several describing models and parameters are taken into consideration and their effects on the performance of the diffusion adaptive networks are examined. The most important parameters that affect the statistical values of the VLC link distribution are the temperature, the salinity level and distance between the nodes [15, 27]. Our contribution in this paper is the presentation of the exact impacts of these parameters on the performance of diffusion networks. With the findings of this paper, one can easily predict the localization, tracking and estimation performance of every diffusion network that is implemented underwater and accordingly, decide about the feasibility of this implantation. The rest of this paper is organized as follows: in part II the diffusion adaptation is formulated for estimating an optimal weight vector. In part III the underwater VLC link model is described, and its parameters are given based on the water characteristics. In part IV, the steady-state performance of the underwater VLC diffusion network is presented. Part V, is for presenting the theoretical and simulation result comparisons. In part VI, we conclude about the feasibility of implementing the diffusion networks using the underwater VLC technology and express our suggestions about the future works.

*Notation:* We used small boldface letters to represent vectors and capital boldface letters for matrixes. The symbol $[.]^*$ denotes complex conjugate for scalars and Hermitian transposition for matrixes. Also, the operator $E[.]$ represents statistical expectation and the notation $\|.\|$ is used for representing the Euclidian norm of a vector. The $bvec(.)$ operator converts block matrices into vectors by stacking the columns of its matrix argument on top of each other.

## II. SYSTEM MODEL

As illustrated in Fig. 1, we consider a diffusion network consists of $N$ active nodes that each contain a processor with the ability to execute adaptive algorithms, the dashed line between the second node and the $k$th node is used to show that several nodes might be existing between them. These nodes can collect data from the surrounding environment like $d_k(i)$ and $u_{k,i}$ (The $k$ and $i$ indices show the nodes and iterations, respectively) and after performing estimation using the adaptive algorithm (which is the Least Mean Square or LMS algorithm in this paper), share the data with other neighbor nodes.

The goal is to estimate the $(M \times 1)$ sized $\boldsymbol{w}^o$ weight vector that connects the $d_k(i)$ and $u_{k,i}$ values with the following linear estimation:

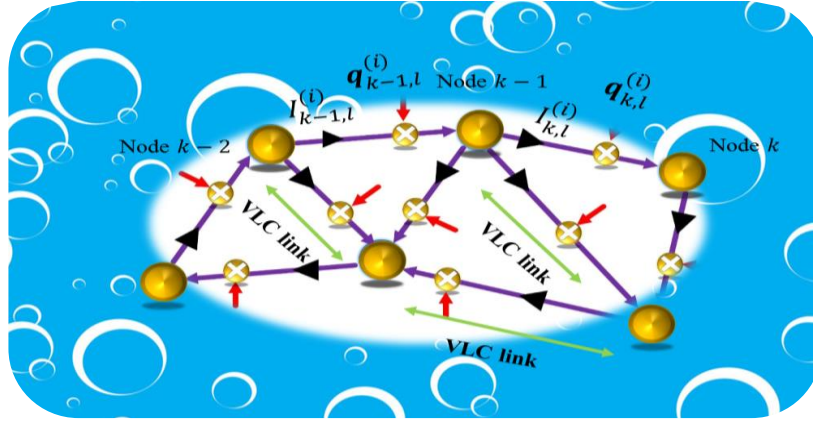

Fig. 2. The diffusion network in underwater conditions using VLC links.

$$d_k(i) = u_{k,i}\mathbf{w}^o + v_k(i) \tag{1}$$

In this relation, the $k$ indices show the node number and the $i$ indices show the iteration number. The convergence to the optimal weight vector will be achieved iteratively by minimizing the following global cost function:

$$J^{glob}(\mathbf{w}) = \sum_{k=1}^{N} E|d_k(i) - u_{k,i}\mathbf{w}^o|^2 \tag{2}$$

The metric to evaluate the performance of the diffusion network is the Mean Square Deviation (MSD) parameter for node $k$:

$$MSD_k \triangleq \lim_{i\to\infty} \mathbb{E}\left[\left\|\widetilde{\boldsymbol{\psi}}_{k-1}^{(i)}\right\|_I^2\right] \tag{3}$$

in this relation $I$ is a $M \times M$ identity matrix and the vector $\widetilde{\boldsymbol{\psi}}_{k-1}^{(i)} = \mathbf{w}^o - \boldsymbol{\psi}_{k-1}^{(i)}$ is the weight error vector. The weighted norm for the exemplary $x$ vector and a Hermitian positive definite matrix $\Sigma > 0$ is defined as: $\|x\|_\Sigma^2 = x^*\Sigma x$. MSD is a means to measure the difference between the optimum weight vector and its network estimation ($\boldsymbol{\psi}_k^{(i)}$) at each node and iteration. The diffusion adaptive network works with two different strategies: the adapt-then-combine (ATC) strategy and the combine then adapt (CTA) strategy that works with low levels of the MSD value in the ideal conditions. However, in non-ideal conditions, the links between nodes are contaminated with channel coefficients and noise. In this paper, we considered that the diffusion network is implemented underwater and with VLC technology. Therefore, the links between the network nodes $k$ and $l$ ($l$ shows the indices of nodes that are connected to node $k$) are known to be contaminated with the optical turbulence coefficients $I_{k,l}^{(i)}$ and noise $q_{k,l}^{(i)}$ and we will have a similar schematic as Fig. 2 for our underwater diffusion network.

The underwater VLC turbulence and noise values follow certain stochastic models that we describe them in part III. Here, we explain the diffusion ATC and CTA strategies in underwater conditions:

*A. Combine then adapt (CTA) diffusion strategy in underwater conditions*

In this strategy first, the nodes combine their communicated data and then perform the local estimation. Through the combination process, the received information from other nodes are added by the Gaussian channel noise ($q_{k,l}^{(i)}$) and multiplied by the VLC link irradiance ($I_{k,l}^{(i)}$), we then have:

$$t_{k,l}^{(i)} = I_{k,l}^{(i)} \psi_k^{(i)} + q_{k,l}^{(i)}, \quad l \in \mathcal{N}_k \tag{4}$$

where $t_{k,l}^{(i)}$ is the received information from node $l$ to node $k$, $I_{k,l}^{(i)}$ is the channel coefficient between these nodes at iteration $i$ and $q_{k,l}^{(i)}$ is the same channel noise with Gaussian distribution and covariance matrices $Q_{k,l} = \mathbb{E}\left[q_{k,l}^{(i)} q_{k,l}^{(i)*}\right]$. With these assumptions the CTA algorithm is given as:

$$\phi_k^{(i-1)} = \sum_{l \in \mathcal{N}_k} c_{k,l} t_{k,l}^{(i-1)} \tag{5}$$

$$\psi_k^{(i)} = \phi_k^{(i-1)} + \mu_k u_{k,i}^* \left(d_k(i) - u_{k,i} \phi_k^{(i-1)}\right) \tag{6}$$

where $c_{l,k}$ are combination coefficients. Choosing different values for these combination values will affect the performance of diffusion algorithms. In our simulations, we used the Uniform policy [14] for determining these coefficients. In this policy we have:

$$c_{l,k} = \begin{cases} \frac{1}{n_k}, & l \in \mathcal{N}_k \\ 0, & otherwise \end{cases} \tag{7}$$

where $n_k \triangleq |\mathcal{N}_k|$ is the size of the neighborhood of node $k$. We can see that all the neighbors of node $k$ are assigned the same weight, $\frac{1}{n_k}$.

*B. Adapt then Combine (ATC) diffusion strategy in underwater conditions*

The second diffusion strategy is ATC where the nodes first perform local estimations and then combine their results. We have:

$$\phi_k^{(i)} = \psi_k^{(i-1)} + \mu_k u_{k,i}^* \left(d_k(i) - u_{k,i} \psi_k^{(i-1)}\right) \tag{8}$$

In order to prevent confusion with CTA algorithm we named the received information, through VLC links, from node $l$ to node $k$ differently and we have:

$$r_{k,l}^{(i)} = I_{k,l}^{(i)} \phi_k^{(i)} + q_{k,l}^{(i)}, \quad l \in \mathcal{N}_k \tag{9}$$

$$\psi_k^{(i)} = \sum_{l \in \mathcal{N}_k} c_{k,l} r_{k,l}^{(i)} \tag{10}$$

The same uniform combination policy for $c_{l,k}$ is considered here. Also, in (9) and (10) $I_{k,l}^{(i)}$ is the channel irradiance and $\boldsymbol{q}_{k,l}^{(i)}$ is the channel noise between nods $l$ and $k$. In the simulations part, we will show that the CTA algorithm works better than the ATC in underwater conditions.

III. UVLC CHANNEL MODEL

The UVLC link properties in this paper are described based on the experimental data for modeling the Log-normal distribution of the link coefficients. UVLC path loss is a function of both attenuation and geometrical losses. The effect of geometrical loss should be considered for LEDs and diffused laser diodes (LDs) [12]. Attenuation is the combination of both absorption and scattering effects. Let $a$ and $b$ denote, respectively, the absorption and scattering coefficients for a given wavelength in a specified water type. The overall attenuation can be then described by the extinction coefficient which is expressed as $c = a + b$. The extinction coefficient takes large values for turbid water (i.e., coastal water and harbor water) while it takes small values in non-turbid water (i.e., pure sea and clear ocean). On the other hand, the geometrical loss is defined as the loss that occurs due to the spreading of the beam between the transmitter and the receiver.

Assuming semi-collimated laser sources, the optical channel coefficient at $k^{th}$ node can be then written as [12]:

$$I_{k,l} \approx D_R^2 \theta_F^{-2} d^{-2} exp(-cD_R^2 \theta_F^{-2} d^{(1-T)}) \tag{11}$$

where $\theta_F$, $D_R$ and $T$ stand for full-width transmitter beam divergence angle, receiver aperture diameter and correction coefficient, respectively. Under the assumption of weak turbulence, the probability density function (PDF) of the turbulence coefficient $I_{k,l}$ in is given by the Log-normal:

$$f_{I_{k,l}}(I_{k,l}) = \frac{1}{I_{k,l}\sqrt{2\pi(4\sigma_{x_{k,l}}^2)}} \exp\left(-\frac{\left(\ln(I_{k,l}) - 2\mu_{x_{k,l}}\right)^2}{2(4\sigma_{x_{k,l}}^2)}\right) \tag{12}$$

where $\mu_x$ and $\sigma_x^2$ denote, respectively, the mean and variance of the log-amplitude coefficient $x_{k,l} = 0.5ln(I_{k,l})$. To ensure that the fading coefficient does not change the value of average power, the fading amplitude is normalized such that $E[I_{k,l}] = 1$, which implies $\mu_{x_{k,l}} = -\sigma_{x_{k,l}}^2/2$. The variance can be written in terms of the scintillation index as $\sigma_{x_{k,l}}^2 = 0.25\ln(1 + \sigma_{I_{k,l}}^2)$ where the scintillation index when laser source with Gaussian beam shape is assumed can be calculated by [19, Eq.7] in conjunction with the power spectrum model of turbulent fluctuations of the sea-water refraction index in [20, Eq. 16].

IV. PERFORMANCE ANALYSIS

In order to examine the exact performance of the diffusion adaptive network in these experimentally modeled link conditions, we present the theoretical analysis of the network performance by considering link coefficient conditions [13]. In this part, the UVLC link coefficients are shown as the $G_i$ matrix in which, each element represents the space-time index of the link coefficient. Finding the impact of the statistical properties of this matrix on the performance of the diffusion network is the main path to analyze the performance theoretically. The theoretical analysis is based on the steady-state error evaluation of the general diffusion LMS algorithm that can be expressed as:

$$\begin{cases} \boldsymbol{\phi}_k^{(i)} = c_{k,i}\boldsymbol{\psi}_k^{(i-1)} + \sum_{l \in \mathcal{N}_k} c_{k,l} I_{k,l}(i)\left(\boldsymbol{\psi}_l^{(i-1)} + \boldsymbol{q}_{k,l}^{(i)}\right) \\ \boldsymbol{\psi}_k^{(i)} = \boldsymbol{\phi}_k^{(i-1)} + \mu_k \boldsymbol{u}_{k,i}^*\left(d_k(i) - \boldsymbol{u}_{k,i}\boldsymbol{\phi}_k^{(i-1)}\right) \end{cases} \quad (13)$$

First we start by defining the following entities [13]:

$$\boldsymbol{\psi}^i \triangleq col\left\{\boldsymbol{\psi}_1^{(i)}, \dots, \boldsymbol{\psi}_N^{(i)}\right\}, \quad \boldsymbol{U}_i \triangleq diag\{\boldsymbol{u}_{1,i}, \dots, \boldsymbol{u}_{N,i}\}$$

$$\boldsymbol{\phi}^i \triangleq col\left\{\boldsymbol{\phi}_1^{(i)}, \dots, \boldsymbol{\phi}_N^{(i)}\right\}, \quad \boldsymbol{d}_i \triangleq col\{d_1(i), \dots, d_N(i)\}$$

$$\boldsymbol{v}_i \triangleq col\{v_1(i), \dots, v_N(i)\}, \boldsymbol{w}^{(o)} \triangleq col\{\boldsymbol{w}^o, \dots, \boldsymbol{w}^o\}$$

$$\boldsymbol{D} \triangleq diag\{\mu_1 \boldsymbol{I}_M, \dots, \mu_N \boldsymbol{I}_M\}, \boldsymbol{q}^i \triangleq col\left\{\boldsymbol{q}_1^{(i)}, \dots, \boldsymbol{q}_N^{(i)}\right\} \quad (14)$$

We assume the $\boldsymbol{q}_{k,l}^{(i)}$s are independent of each other:

$$\boldsymbol{Q} \triangleq \mathbb{E}[\boldsymbol{q}^i(\boldsymbol{q}^i)^*] = diag\{\boldsymbol{Q}_1, \dots, \boldsymbol{Q}_N\} \quad (15)$$

By considering $d_k(i) = \boldsymbol{u}_{k,i}\boldsymbol{w}^o + v_k(i)$ we have:

$$\boldsymbol{d}_i = \boldsymbol{U}_i \boldsymbol{w}^{(o)} + \boldsymbol{v}_i \quad (16)$$

using these definitions (13) changes to:

$$\boldsymbol{\phi}^{i-1} = \boldsymbol{G}_i \boldsymbol{\psi}^{i-1} + \boldsymbol{q}^{i-1}$$

$$\boldsymbol{\psi}^i = \boldsymbol{\phi}^{i-1} + \boldsymbol{D}\boldsymbol{U}_i^*(\boldsymbol{d}_i - \boldsymbol{U}_i \boldsymbol{\phi}^{i-1}) \quad (17)$$

where $\boldsymbol{G}_i$ is the link turbulence coefficient matrix:

$$\boldsymbol{G}_i = \begin{bmatrix} c_{11} & I_{12}(i)c_{12} & \cdots & I_{1N}c_{1N} \\ \vdots & c_{22} & \cdots & \vdots \\ \vdots & \vdots & \cdots & \vdots \\ I_{N1}(i)c_{N1} & \cdots & \cdots & c_{NN} \end{bmatrix} \quad (18)$$

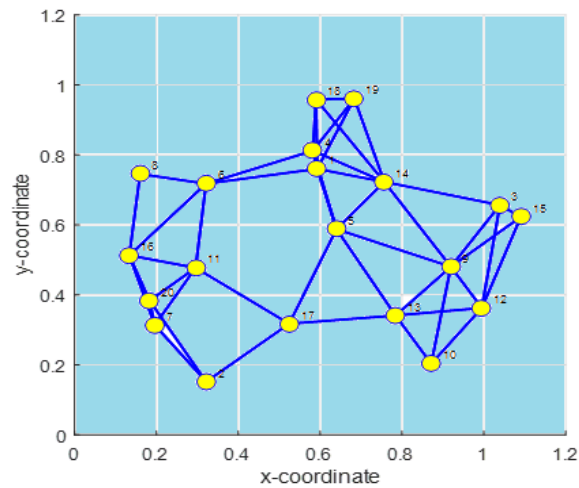

Fig. 3. The topology of a diffusion network implemented underwater.

TABEL I. Channel turbulence coefficient log-amplitude variances for different distances

| $d(m)$ | 1 | 2 | 3 | 4 | 5 |
|---|---|---|---|---|---|
| $\sigma^2_{x_{k,l}}$ | $1.07 \times 10^{-3}$ | $3.5 \times 10^{-3}$ | $6.97 \times 10^{-3}$ | $1.13 \times 10^{-2}$ | $1.64 \times 10^{-2}$ |
| $d(m)$ | 6 | 7 | 8 | 9 | 10 |
| $\sigma^2_{x_{k,l}}$ | $2.22 \times 10^{-2}$ | $2.85 \times 10^{-2}$ | $3.54 \times 10^{-2}$ | $4.27 \times 10^{-2}$ | $5.04 \times 10^{-2}$ |
| $d(m)$ | 11 | 12 | 13 | 14 | 15 |
| $\sigma^2_x$ | $5.84 \times 10^{-2}$ | $6.67 \times 10^{-2}$ | $7.52 \times 10^{-2}$ | $8.39 \times 10^{-2}$ | $9.28 \times 10^{-2}$ |
| $d(m)$ | 16 | 17 | 18 | 19 | 20 |
| $\sigma^2_{x_{k,l}}$ | $1.02 \times 10^{-1}$ | $1.11 \times 10^{-1}$ | $1.2 \times 10^{-1}$ | $1.29 \times 10^{-1}$ | $1.38 \times 10^{-1}$ |

These coefficients have a direct impact on the performance of the diffusion adaptive network. In [13] the MSD of the diffusion LMS algorithm for the CTA strategy is expressed as:

$$MSD = \frac{1}{N} \boldsymbol{g}(\boldsymbol{I} - \overline{\boldsymbol{F}})^{-1} \boldsymbol{r} \qquad (19)$$

where $\boldsymbol{r} = bvec\{\boldsymbol{I}_{NM}\}$. For each node in steady-state condition we have:

$$MSD_k = \boldsymbol{g}(\boldsymbol{I} - \overline{\boldsymbol{F}})^{-1} bvec\{\boldsymbol{J}_{r,k}\} \qquad (20)$$

where:

$$\boldsymbol{J}_{r,k} = diag\{\boldsymbol{0}_{(k-1)M}, \boldsymbol{I}_M, \boldsymbol{0}_{(N-k)M}\} \qquad (21)$$

In [13, Eq.51] and [13, Eq.57] it has been shown respectively, that the $\overline{\boldsymbol{F}}$ and $\boldsymbol{g}$ are directly dependent to the $\boldsymbol{G}_i$ matrix and therefore, if we replace the UVLC link coefficients with the FSO link coefficients in [13], the $MSD_k$ depends on the UVLC link irradiance coefficients.

TABEL II. Channel turbulence coefficient

| T (°C) | S (PPT) | $\sigma^2_{x_{k,l}}$ |
|---|---|---|
| 1 | 35 | $8.04 \times 10^{-5}$ |
| 28 | 35 | $1.57 \times 10^{-3}$ |
| 20 | 33 | $1.04 \times 10^{-3}$ |
| 20 | 36.5 | $1.09 \times 10^{-3}$ |

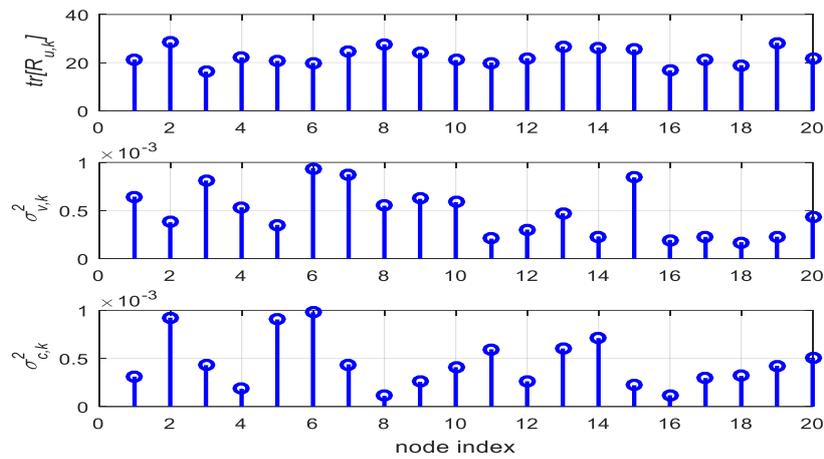

Fig. 4. The typical network parameters for our simulation setup.

## V   SIMULATION RESULTS

In this section, we consider performance results of an underwater diffusion network with VLC links. The network is assumed to have $N = 20$ nodes and it is depicted in Fig. 3. The topology of the diffusion network is adopted from [20].

Unless otherwise stated, we consider receiver aperture diameter of $D_R = 5\ cm$, full width transmitter beam divergence angle of $\theta_F = 6°$ and total transmit power of $P_T = 1$ W. Assuming clear ocean, the extinction and correction coefficients are given, respectively, as $c = 0.15$ and $T = 0.05$ [15]. We further calculate the scintillation index ($\sigma^2_{I_{k,l}}$) based on [25, Eq.7] in conjunction with [26, Eq.16] assuming salinity of 35 PPT and temperature of 20°C. Utilizing the computed $\sigma^2_{x_{k,l}}$[1], we calculate log-amplitude variance ($\sigma^2_{x_{k,l}}$). We consider the fact that nodes may have different link distances and consider the range of $d = 1, 2, \ldots, 20\ m$. The corresponding values of log-amplitude variances are listed in Table I. Also, for different salinity and temprature levels of the water, the log-amplitude variances are given in TABLE II.

---

[1] For computing log-amplitude variance, dissipation rate of mean-squared temperature of $1 \times 10^{-3} K^2 s^{-3}$ and dissipation rate of turbulent kinetic energy per unit mass of fluid is assume as $1 \times 10^{-2} m^2 s^{-3}$ are assumed (These are parameters of spectrum model in [26]).

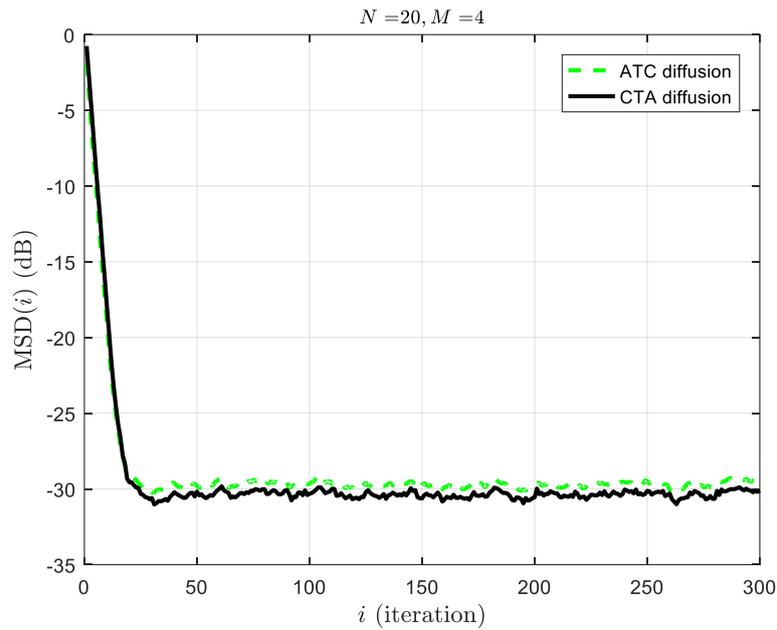

Fig. 5. The Underwater performance of the diffusion network for the temperature 1 °C and salinity level of 35 PPT.

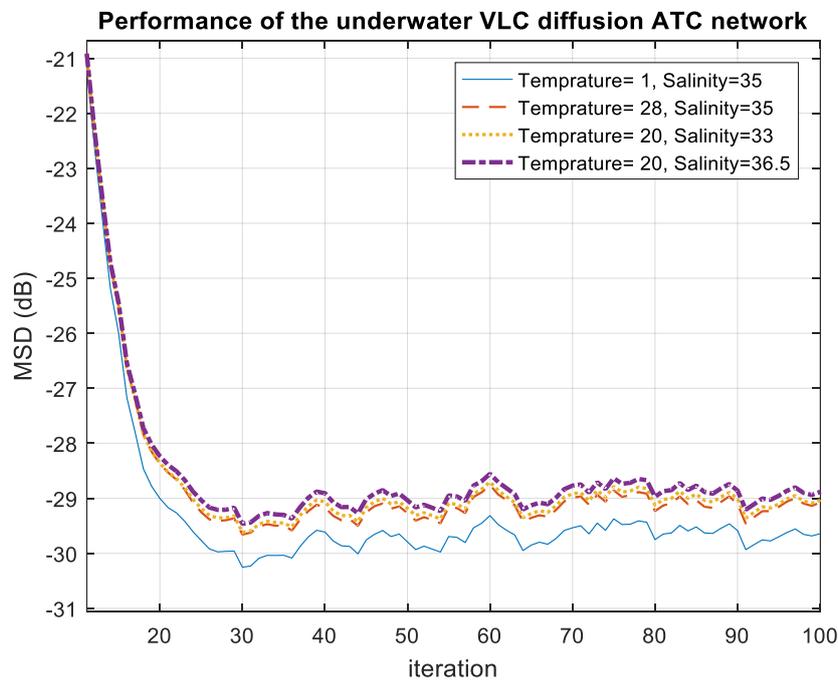

Fig. 6. The performance of the diffusion ATC algorithm at node 1 from iteration 10 to 100 for various temperature and salinity levels.

These links are considered to be contaminated with the Gaussian noise with the variance $\sigma_{c,k}^2$. and the Log-normally distributed turbulence coefficients with the Log-variance of $\sigma_x^2$. The measurement noise variance is also $\sigma_{v,k}^2$ and all of these parameters along with the input covariance matrix traces are given in Fig. 4. The network is used to estimate a vector with the size of $4 \times 1$ and with entries of $\boldsymbol{w}^o = [1\ 1\ 1\ 1]^T / \sqrt{4}$.

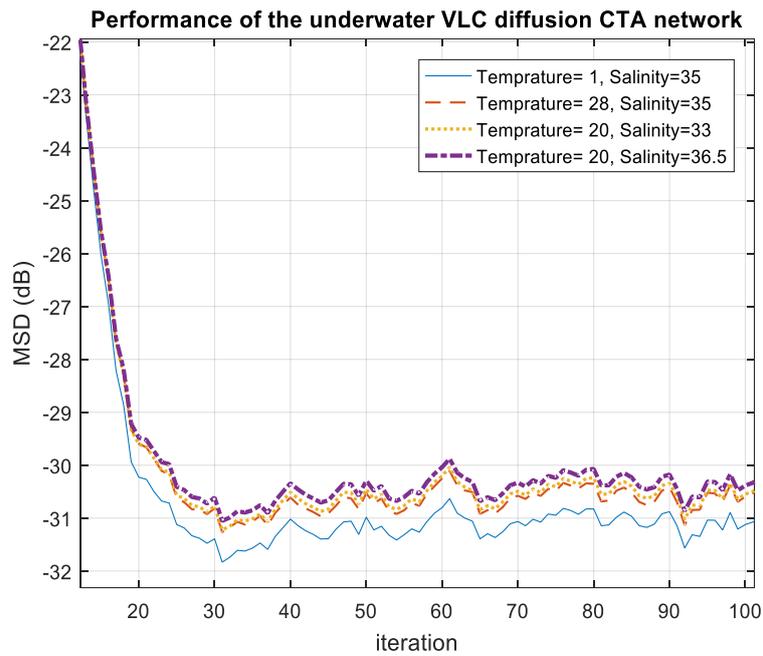

Fig. 7. The performance of the diffusion CTA algorithm at node 1 from iteration 10 to 100 for various temperature and salinity levels.

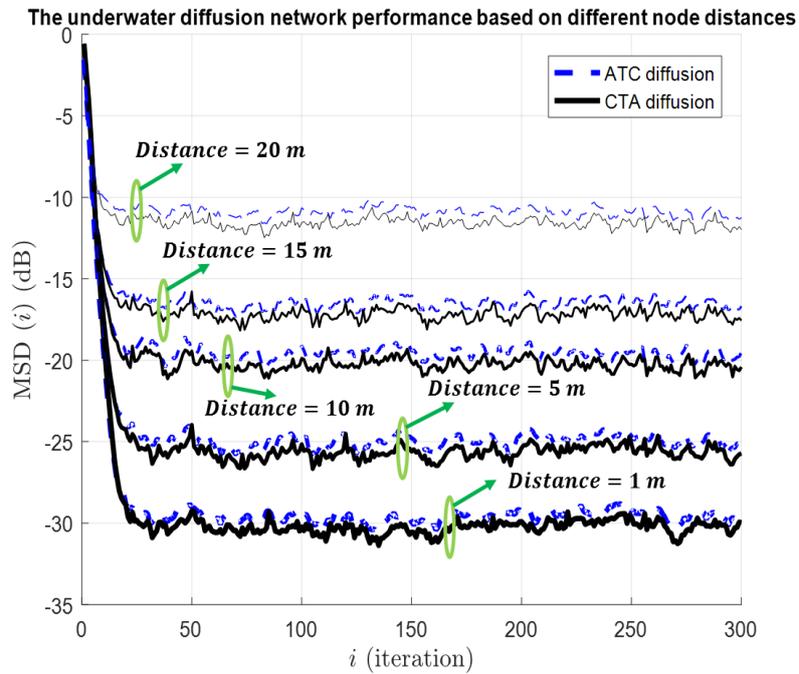

Fig. 8. The effects of various distances between the diffusion network nodes on the underwater performance of the diffusion network. It is important to mention that the CTA results are better than the ATC ones.

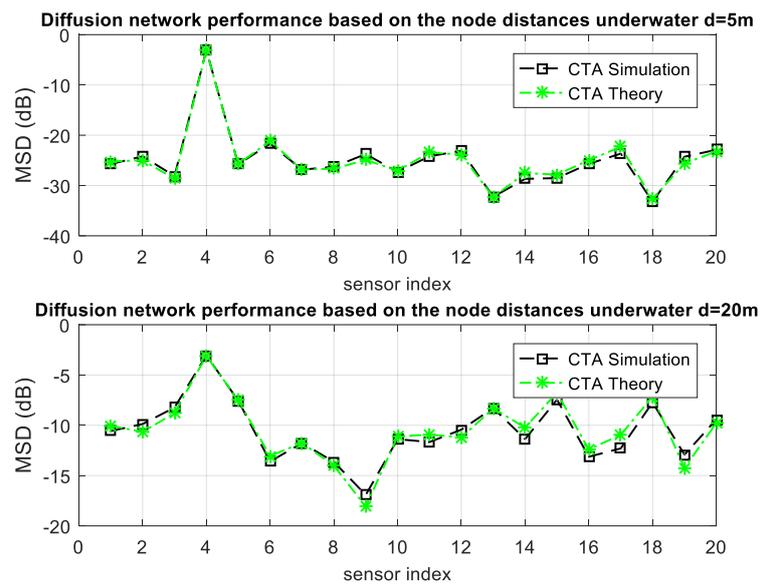

Fig. 9. The theoretical vs. simulation performance of the diffusion network for the different node distances.

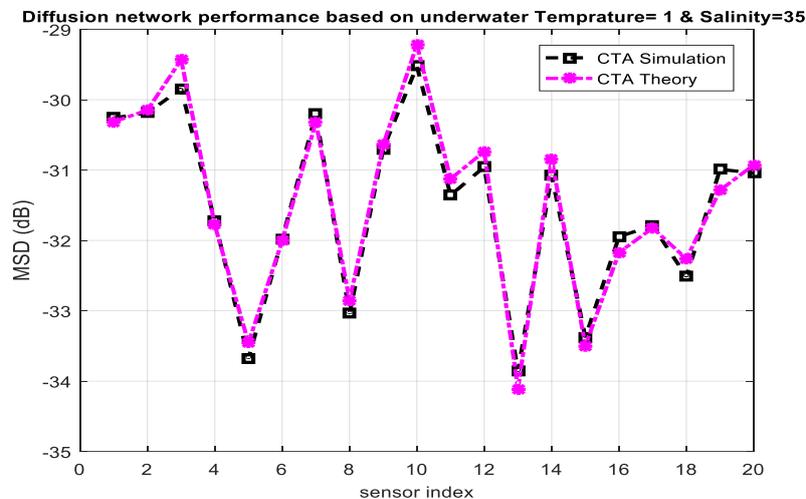

Fig. 10. The theoretical vs. simulation performance of the diffusion network for underwater temperature 1 ℃ and salinity level of 35 PPT.

In our first simulation, we considered both the ATC and CTA diffusion algorithms in the underwater VLC condition. The results are depicted in Fig. 5 and they show that the diffusion network can converge to the optimal weight vector with a reasonable error level. This means that the diffusion network can converge to the estimation parameters if it is implemented underwater using the VLC technology. For this simulation, we only considered the case in which the temperature of the water is 1 ℃ and the salinity level is 35 PPT.

For this simulation, we considered that the distances between the network nodes to be 1 meter and as we expected, the performance of the CTA strategy is better than the ATC in underwater conditions. In our second simulation, we only considered the results of the ATC scheme. In Fig. 6 we can see the effects of the different water temperature and salinity levels (based on the data in Table 1) on the performance of the ATC diffusion adaptive networks.

Next, we consider the performance of the CTA network in the underwater condition. As we can see in Fig. 7, the effects of the VLC links become more serious as the turbulence variance increase. However, we use these simulations to acquire the exact error values of the diffusion networks in the UVLC conditions. For example, the error values of the ATC strategy in the temperature 1 ℃ and salinity level of 35 PPT is over -30 dBs while for the CTA it is over -32 dBs.

The results in all of these simulations showed that even a small difference in the amount of salt in water and the temperature can affect the performance of the underwater implemented an adaptive network. Also, it is important to mention that in all cases, the error levels of the CTA diffusion strategy are lower than the ones belonging to the ATC strategy.

In our next simulations, we examined the effects of the distances between the nodes in the network performance. For simulations of this part, the turbulences that were induced based on the salinity and temperature levels of the water, are neglected to show the exact impact of the distances in Fig. 8.

The results in Fig. 8, are the first performance analysis of the diffusion networks that consider the effects of distance between the nodes in the performance results. The diffusion network can converge in even very long distances between the nodes. However, as the distance becomes higher, the turbulence induced fading becomes stronger and the transmitted data for the receiver node becomes more faded. This accordingly causes the network MSD error to rise based on the distance range.

Along with the simulation results that comprised of both transient and steady-state performances of the adaptive network [14], here we compare the stedy-state theoretical and simulation performances of the diffusion network in the UVLC conditions. It is important to mention that as we only considered the steady-state performance of the CTA strategy, the theoretical and simulation results comparisons are only presented for this strategy. The CTA algorithm performs slightly better than the ATC network and presentation of the theoretical calculations of both CTA and ATC is not common in the papers [13]. First, we consider the distance effects on this network performance in Fig. 9.

As we can see in the 5 meter node distance condition, the error performance is around -25 dB and when the distances between the nodes become 20 meters, the error performance degrades and becomes around -10 dB. Next, we consider the effects of the salinity and temperature values on the theoretical and simulation performance of the diffusion network in Fig. 10.
The conformity of the theoretical and simulation results showed the accuracy of our calculations and the precise effects of the statistical properties of the UVLC link coefficients on the performance of the diffusion networks.

V. CONCLUSION

In this paper, we presented the theoretical and simulation performance results of an adaptive diffusion network that is implemented underwater with the usage of the VLC technology. The VLC links between the nodes are considered to follow the Log-normal distributions with the variances that are related to the temperature and salinity level of the examined water and the distances between the nodes of the diffusion network. We showed that as these parameters get higher, the turbulence of the VLC link increases and the performance of the diffusion network degrades. However, as the overall turbulence level that is modeled with the Log-normal distribution is weak, the performance of the diffusion network is acceptable. Therefore, the underwater implementation of the diffusion networks, especially with the CTA strategy, is recommended for various applications. In future works, we will investigate the effects of other more realistic underwater turbulence models on the performances of diffusion networks and by this, we will come closer to the implementing sensor network under seas for real-world applications.

AKNOWLEDGMENT

The authors would like to thank Mr. Mohammed Elamassie from Özyeğin University, Istanbul, Turkey, for sharing their experimental data about modeling the underwater VLC links.